\documentclass[amsmath,amssymb,aps,prc,twocolumn,superscriptaddress,showpacs,preprintnumbers,nofootinbib]{revtex4-1}

\def\fsize{8.5}
\def\fsizeb{8.8}

\usepackage{graphicx,color}
\usepackage{multirow}
\usepackage{times}
\usepackage{bm}
\usepackage{epstopdf}
\usepackage{enumerate}

\usepackage[normalem]{ulem}  
\newcommand{\comment}[1]{}
\renewcommand\sout{\bgroup \color{red} \ULdepth=-.5ex \ULset}

\begin{document}

\title{
Momentum-dependent potential and collective flows
within 
the relativistic quantum molecular dynamics approach
based on relativistic mean-field theory
}

\author{Yasushi Nara}
\affiliation{
Akita International University, Yuwa, Akita-city 010-1292, Japan}
\affiliation{Frankfurt Institute for Advanced Studies, 
D-60438 Frankfurt am Main, Germany}

\author{Tomoyuki Maruyama}
\affiliation{College of Bioresource Sciences, Nihon University, 
Fujisawa 252-0880, Japan}

\author{Horst Stoecker}
\affiliation{Frankfurt Institute for Advanced Studies, 
D-60438 Frankfurt am Main, Germany}
\affiliation{Institut f\"ur Theoretische Physik,
 Johann Wolfgang Goethe Universit\"at, D-60438 Frankfurt am Main, Germany}
\affiliation{GSI Helmholtzzentrum f\"ur Schwerionenforschung GmbH, D-64291
Darmstadt, Germany}

\date{\today}
\pacs{
25.75.-q, 
25.75.Ld, 
25.75.Nq, 
21.65.+f 
}

\begin{abstract}
Relativistic quantum molecular dynamics
based on the relativistic mean field theory (RQMD.RMF)
is extended by including momentum-dependent potential.
The equation of state (EoS) dependence of
the directed and the elliptic flow of protons in the beam energy range of
$2.3 < \sqrt{s_{NN}}< 20$ GeV is examined.
It is found that the directed flow
depends strongly on the optical potential at high energies,
$\sqrt{s_{NN}} > 3 $ GeV, where no information is available experimentally.
The correlation between effective mass at saturation density
and the optical potential is found:
smaller values of effective mass require
smaller strengths of the optical potential
to describe the directed flow data.
This correlation can also be seen in the beam energy dependence of
the elliptic flow at $\sqrt{s_{NN}}>3$ GeV,
although its effect is rather weak.
On the other hand, stiff EoS is required to describe
the elliptic flow at lower energies.
Experimental constraints on the optical potential from $pA$ collisions
will provide important information on the EoS at high energies.
The proton directed and the elliptic flow are well described in
the RQMD.RMF model from $\sqrt{s_{NN}}=2.3$ to 8.8 GeV.
In contrast,
to reproduce 
the collapse of the directed flow above 10 GeV,
pressure has to be reduced, which 
indicates
a softening of the EoS
around $\sqrt{s_{NN}} =10 $ GeV.
\end{abstract}

\maketitle

\section{Introduction}

High energy heavy-ion collisions
provide a unique opportunity to explore the properties of strongly
interacting QCD matter for a wide range of temperatures and densities.
In particular, collisions in the energy range of $2 < \sqrt{s_{NN}} < 20$ GeV
create high baryon density matter, and
should be the best place to search for
the onset of a phase transition as well as critical point
in QCD matter.
The ongoing beam energy scan program (BES)~\cite{Singha:2016mna,Adam:2019wnb}
at the BNL-RHIC-STAR- and
the CERN-SPS-NA49 and -NA61/SHINE experiments~\cite{Turko:2018kvt}
have measured beam energy, collision system, and centrality dependence
of observables 
such as collective flows, fluctuations of conserved charges,
which are expected to be sensitive to a phase transition 
and/or critical point. 
Future experiments, such as RHIC-BESII~\cite{BESII}, STAR FXT, CBM and HADES at
FAIR~\cite{FAIR,Sturm:2010yit}, BM@N and MPD at NICA~\cite{NICA},
HIAF at Canton, as well as the proposed J-PARC-HI~\cite{HSakoNPA2016},
will further offer excellent opportunity
to
explore the highest density baryonic matter sector of QCD, 
and determine the phase structure of QCD
with high statistics data.

To extract the information on the properties of high dense
QCD matter from heavy-ion experimental data,
details of the collision dynamics have to be understood.
For this purpose, the transport models such as
non-equilibrium microscopic transport models%
~\cite{RQMD1989,Sorge:1995dp,UrQMD1,UrQMD2,HSD1999,GiBUU},
hydrodynamical models~\cite{Ivanov:2019gxm},
and hybrid models~\cite{Petersen:2008dd,Karpenko:2015xea,Batyuk:2016qmb,Shen:2017bsr,%
Denicol:2018wdp,Akamatsu:2018olk,Shen:2020}
have been used
to simulate space-time evolutions of 
hot and dense matter created in nuclear collisions at high baryon density
regions.
It has been argued for a long time
the determination of
the equation of state (EoS)
from collective flows such as
directed as well as elliptic flow, as they
are sensitive to the EoS
~\cite{Stoecker:1980vf,Stoecker:1981pg,Buchwald:1984cp,%
Stoecker:1986ci,Hartnack:1994ce,
Ollitrault:1992,Danielewicz:2002pu,Stoecker:2004qu}.
For instance, fluid dynamics predicts 
negative directed flow of protons
at the vicinity of the softest point
in the EoS with a first-order phase transition%
~\cite{Rischke:1995pe,Brachmann:1999xt,Csernai:1999nf,Ivanov:2014ioa,Ivanov:2016sqy},
which is also confirmed by microscopic transport model calculations%
~\cite{Li:1998ze,Nara:2016phs,Nara:2016hbg}.
The NA49~\cite{NA49prc} and the STAR~\cite{STARv1,STARv1new} collaborations
discovered the negative proton directed flow
at $\sqrt{s_{NN}} > 8$ GeV,  which locates
rather higher beam energies than the AGS energies that
is expected to be a softest point by
most of the theoretical predictions.

The time evolution of heavy-ion collisions generally consist of
far from an equilibrium state to
late possible equilibrium stage followed by a freeze-out process.
Non-equilibrium microscopic transport approach is
a theoretical framework
to simulate a collision of nuclei from initial to final stages in a unified way,
and it has been widely used to describe nuclear collisions from low to
high energies,
see Ref~\cite{ModelComp}
for the recent comparison of heavy-ion transport codes.
A relativistic transport approach
based on the relativistic mean-field theory of Walecka type,
called relativistic Boltzmann-Uehling-Uhlenbeck (RBUU)
has been formulated, and several transport codes have been
developed for heavy-ion collisions%
~\cite{RBUU,GiBUU,RVUU,Li:1989zza,Cassing:1991cy,RLV}.
On the other hand, the quantum molecular dynamics (QMD)
~\cite{Aichelin:1986wa,Aichelin:2019tnk,UrQMD1,UrQMD2}
is an $N$-body approach, which simulates multi-particle collision dynamics
beyond the time evolution of one-particle distribution function
like RBUU models.
Therefore, the QMD model
can be applied 
to study, for example,
multi-fragmentations
and event-by-event fluctuations.
The relativistic version of the QMD model (RQMD) has been developed
by the Lorentz scalar treatment of the Skyrme potential%
~\cite{RQMD1989,Sorge:1995dp,Maruyama:1991bp,Maruyama:1996rn,
Mancusi:2009zz,Isse:2005nk}.
RQMD with the relativistic mean-field has been developed
in Ref.~\cite{Fuchs:1996uv} for
the intermediate energy heavy-ion collisions
up to $E_\mathrm{lab}= 2A$ GeV.
Recently, RQMD based on
the relativistic mean-field theory (RQMD.RMF)~\cite{Nara:2019qfd}
has been implemented into the transport code JAM~\cite{JAMorg}
to simulate high energy nuclear collisions.
It is shown that this relativistic transport approach RQMD.RMF reproduces
the beam energy dependence of the directed as well as the elliptic flow
from $\sqrt{s_{NN}}=2.4$ GeV up to $8$ GeV.

The importance of 
the momentum dependence of the mean-field 
was realized for the extraction of the EoS
from heavy-ion collisions~\cite{Aichelin:1987ti,Welke:1988zz}.
An extension of the RBUU model by introducing an additional
momentum-dependent potential
has been formulated in Ref.~\cite{Weber:1992qc}, which remedies
the problem of too repulsive potential in 
the Walecka model.
Numerical simulations of RBUU with momentum-dependent interaction
was performed at low and intermediate heavy-ion collisions
at $E_\mathrm{lab}<2 A$ GeV within polynomial approximation
for the momentum-dependent potentials~\cite{Maruyama:1992vx,Maruyama:1993jb}.
RBUU approach with momentum-dependent scalar and vector form factor
was applied for the study of the beam energy dependence of
the directed and the elliptic flow~\cite{Sahu:1998vz}.

RQMD approach with relativistic mean-field including explicit momentum-dependent
interactions have not been developed to date.
In this paper, we extend our RQMD.RMF approach~\cite{Nara:2019qfd}
by incorporating momentum-dependent potential
in line with Ref.~\cite{Weber:1992qc},
and apply it to high energy heavy-ion collisions
at 
$2.3 \lesssim \sqrt{s_{NN}} \lesssim 20$ GeV
 ($1 \lesssim E_\mathrm{lab} \lesssim 160 A$ GeV)
to investigate the effects of momentum-dependence 
on the collective flow.
Optical potential has been extracted by experiments up to
the beam energy of $E_\mathrm{lab}=1$ GeV. Therefore, we
test different strengths of the optical potential at $E_\mathrm{lab}>1$ GeV
to study the sensitivity of the flows to the optical potential.

This paper is organized as follows:
Section~\ref{sec:model} describes the nonlinear $\sigma$-$\omega$ model
with momentum-dependent potential
and its implementation into the RQMD framework.
Section~\ref{sec:results} presents the results
for the beam energy dependence of the directed
and the elliptic flow, as well as the rapidity dependence of the directed flow.
The summary is given in Sec.~\ref{sec:summary}.

\section{Model}
\label{sec:model}

We first present our EoS from relativistic mean-field theory
with momentum-dependent potentials, and then describe
how to implement it into the framework of RQMD approach.
This approach has been realized in the transport code JAM,
which enables us to simulate nuclear collisions at high energies.

\subsection{EoS from the relativistic mean-field with momentum-dependent
potential}

A covariant treatment of the momentum dependence of the relativistic
potentials
in the relativistic mean-field theory
was formulated in Ref.~\cite{Weber:1992qc}.
Here we shall employ a Lorentzian form of
momentum-dependent potential
which depends only on the spacial part of momentum
neglecting the energy dependence, which is related to
nonlocality in time.
This is consistent with our assumption of time-fixation conditions
specified below in our RQMD approach.
Thus,
we introduce the following momentum-dependent scalar and vector potential:
\begin{align}
  V_{s}^\mathrm{MD} &= \frac{\bar{g}_s^2}{m_s^2}
       \int d^3p\frac{m^{*}}{p^{*}_0}
       \frac{f(x,p)}{1+(\bm{p}-\bm{p}')^2/\Lambda_s^2}\,, \\
  V_{\mu}^\mathrm{MD} &= \frac{\bar{g}_v^2}{m_v^2}
       \int d^3p\frac{p^*_\mu}{p^{*}_0}
       \frac{f(x,p)}{1+(\bm{p}-\bm{p}')^2/\Lambda_v^2} \,,
  \label{eq:md}
\end{align}
where $f(x,p)$ is a phase space distribution function.
At zero temperature, it is given by
\begin{equation}
 f(x,p) = \frac{g_N}{(2\pi)^3}\theta(p_F- |\bm{p}|).
\end{equation}
where $p_F$ is the Fermi momentum, and  $g_N=4$ 
is the degeneracy factor for spin and isospin of nucleons.
In the actual implementation into the RQMD model, the arguments
of the momentum-dependent potentials are replaced by the
relative momentum in the two-body center-of-mass
frame between interacting particles to maintain the covariance
of the theory.

The energy density for nuclear matter in the relativistic mean-field theory
with $\sigma$- and $\omega$- meson-baryon interactions
with momentum-dependent potentials is given by~\cite{Weber:1992qc}
\begin{align}
e &= \int d^3p\, p_{0}f(p) + U(\sigma) \nonumber\\
  &+ \frac{1}{2}\int \frac{d^3p}{p^*_0}\left(
   m^*V_s^\mathrm{MD}- p^{*\mu}V_\mu
    \right)f(p).
\end{align}
Here,
the vacuum mass $m$ and canonical momentum $p_\mu$
are modified by the scalar potential $S$ and the vector
potential $V_\mu$, which define
the effective mass $m^*$ and kinetic momentum $p^*$, respectively:
\begin{align}
m^* &= m - S = m - g_s\sigma - V_s^\mathrm{MD},\\
p_\mu^* &= p_\mu - V_\mu = p_\mu - g_v \omega_\mu - V^\mathrm{MD}_\mu\,.
\end{align}
The mass-shell constraint $p^{*2}-m^{*2}=0$
is consistent with
the single-particle energy as
\begin{equation}
p_0=\sqrt{m^{*2} + \bm{p}^{*2}}+ g_v\omega_0 + V_0^\mathrm{MD}\,.
\end{equation}
For the scalar field,
the following nonlinear self-interaction is introduced~\cite{Boguta}:
\begin{equation}
  U(\sigma) = \frac{m_\sigma^2}{2}\sigma^2
         + \frac{g_2}{3}\sigma^3
         + \frac{g_3}{4}\sigma^4 \,.
\label{eq:sigmapot}
\end{equation}
The $\sigma$ and $\omega$ field are obtained
by solving the self-consistent equations
\begin{equation}
  m_s^2 \sigma + g_2\sigma^2 + g_3\sigma^3=g_s\rho_s \,,~~
  m_v^2 \omega^0 = g_v \rho_v\,.
  \label{eq:sigma}
\end{equation}
Here $\rho_s=\int d^3p \frac{m^*}{p^*_0}f(p)$ is the scalar density,
and $\rho_v=\int d^3p f(p)$ is a zeroth component of the vector density.

In  order to fix the parameters in the momentum-dependent potentials,
we fit the real part of the experimentally determined
nucleon-nucleus optical potential~\cite{Hama:1990vr}
together with the binding energy per nucleon
$ E/A  = p_0(p_F) - m_N= 16~\mathrm{MeV}$.
We define the optical potential by subtracting kinetic
energy from the single-particle energy of nucleon%
~\cite{Feldmeier:1991ey,Danielewicz:1999zn,Kapusta:2006pm}
\begin{equation}
 U_\mathrm{opt}(p)= p_0(p) - \sqrt{m_N^2+\bm{p}^2},
 \label{eq:uopt}
\end{equation}
where $m_N$ is the free nucleon mass.
This optical potential is similar to the Schr\"odinger-equivalent
potential at low to moderate momenta, but it approaches
a constant value at high-energy limit
in contrast to the Schr\"odinger-equivalent potential
that linearly depends on the kinetic energy for nonzero vector potential.
Remaining parameters of the EoS are determined 
by the condition that ground state is a minimum in the EoS
at the  normal nuclear matter density $\rho_B=\rho_0=0.168$ 1/fm$^3$:
$P = \rho_B^2\partial(e/\rho_B)/\partial\rho_B|_{\rho_B=\rho_0}=0$
for a given incompressibility
$K = 9\rho_B^2\partial^2(e/\rho_B)/\partial\rho_B^2|_{\rho_B=\rho_0}$
and effective nucleon mass $m^*(\rho_0)$ 
at the normal nuclear matter density.
The parameter sets are given in Table \ref{table:eos}
for different incompressibilities, effective masses,
and optical potentials
to investigate the influence of EoS on the collective flows.

\begin{table}
\caption{Parameters for the relativistic mean-field
theory with nonlinear scalar interaction 
and momentum-dependent potentials.
A binding energy of $E/A=-16$ MeV at normal nuclear matter density of
$\rho_0=0.168$ 1/fm$^3$,
a $\sigma$ mass of $m_s=0.55$ GeV,
and an $\omega$ mass of $m_v=0.783$ GeV are used.
}
%

\begin{tabular}{ccccccccc}\hline\hline
       &       &  NS1  & NS2      & NS3    & MD1   & MD2   & MD3 & MD4 \\\hline
$K$    & (MeV) & 380   & 210      & 380    & 380   & 380   & 380 & 210 \\
$m^*/m$ &      & 0.83  & 0.83     & 0.7    & 0.65  & 0.65  & 0.65 & 0.83\\
$U_\mathrm{opt}(\infty)$
       & (MeV) &  95   &  98      & 200   & 95     & 30    & $-0.4$& 67 \\
$g_s$  &       & 6.448 & 7.902    & 8.864 & 9.030  & 9.233 &5.439  & 4.059 \\
$g_v$  &       & 6.859 & 6.859    & 10.07 & 6.740  & 3.888 & 0.0   &5.632 \\
$g_2$  & (1/fm)& $-38.0$&44.31  & 2.191 & 4.218    &4.012  &$-15.59$&$-160.3$ \\
$g_3$  &       & 339.6 & 21.99  & 27.07 & 6.667    &5.520  & 391.9  &2684 \\
$\bar{g}_s$&    & -        & -     &  -    & 3.186 &2.502  & 7.711  &5.544  \\
$\bar{g}_v$&    & -        & -     &  -    & 8.896 &10.43  & 11.22  &3.926 \\
$\Lambda_s$&(GeV)& -        & -     &  -   & 0.641 &0.4897 & 1.702  &0.704 \\
$\Lambda_v$&(GeV)& -        & -     &  -   & 1.841 &2.489  & 1.898  &4.252 \\

\hline\hline
\end{tabular}
\label{table:eos}
\end{table}

\begin{figure}[tbh]
\includegraphics[width=\fsize cm]{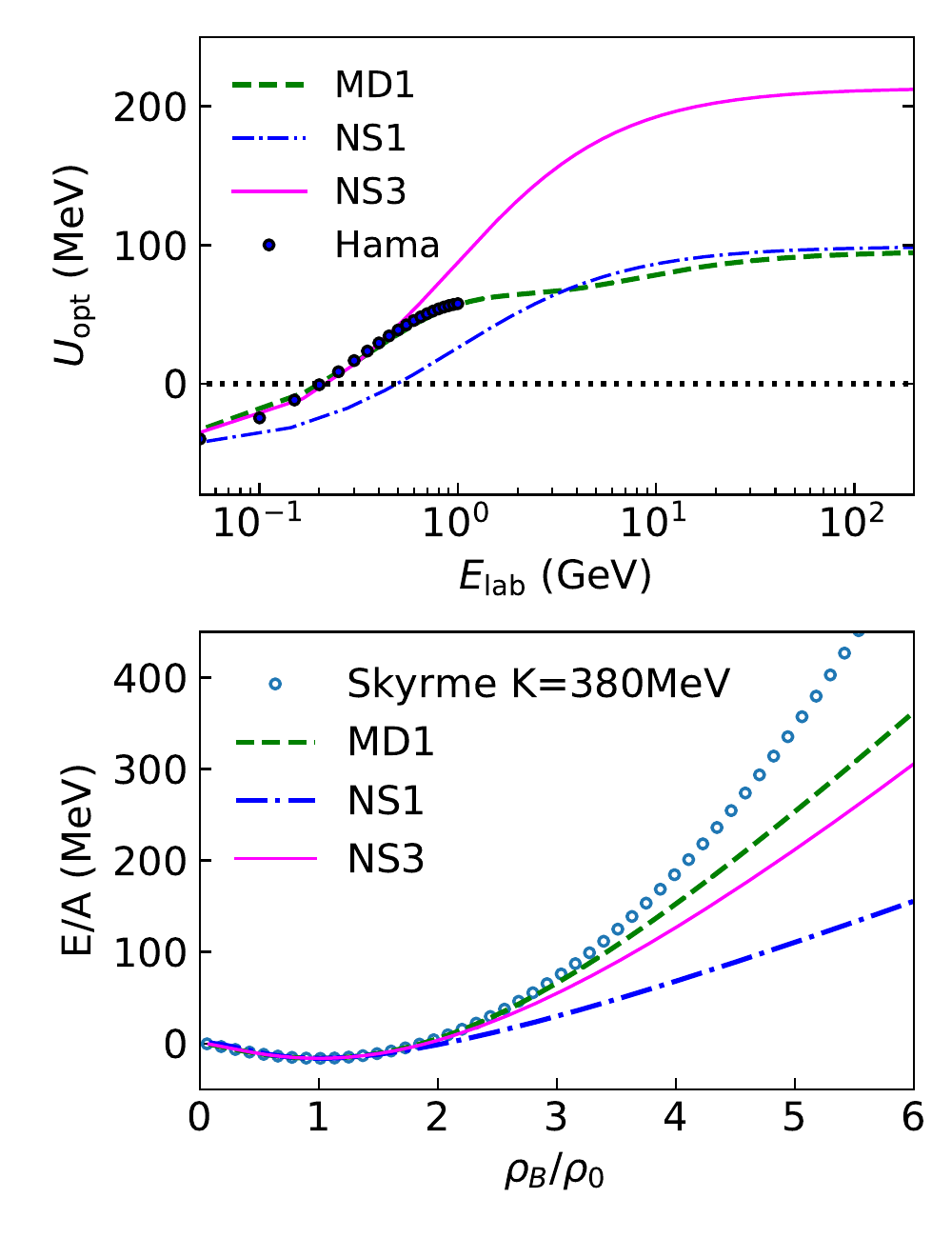}
\caption{Upper panel: optical potentials in normal nuclear matter density
as a function of incident energy.
Lower panel: total energy per nucleon as a function of the normalized
baryon density at zero temperature.
The dashed, dot-dashed, and solid lines
show the results for
the parameter set MD1, NS1, and NS3, respectively.
The hard EoS from Skyrme potential with $K=380$ MeV is shown
by open circles.
The full circles correspond to the results
of the global Dirac optical model fit to
$p$-nucleus elastic scattering data
by Hama \textit{et al}.~\cite{Hama:1990vr}.
}
\label{fig:eosp1}
\end{figure}

In the upper panel of Fig.~\ref{fig:eosp1},
we compare energy dependence of
the optical potential defined by Eq.~(\ref{eq:uopt})
at normal nuclear matter density
with the parameters with and without momentum-dependence.
The parameter set NS3 ($m^*/m=0.7$) reproduces
the experimentally determined optical potential~\cite{Hama:1990vr}
up to $E_\mathrm{lab}=0.5$ GeV,
while NS1 ($m^*/m=0.83$) significantly underestimates the data.
In contrast, the optical potential of NS3 has much higher values
than the data at higher beam energies above $E_\mathrm{lab}=1$ GeV,
as is well known that the Walecka type model has strong 
energy dependence.
Analysis of the directed flow data by the transport models
with $\sigma$-$\omega$ interactions found that
the parameter set with $m^*/m \approx 0.7$ 
fits the flow data at lower beam energies $E_\mathrm{lab}<0.4 A$ GeV,
while the parameter set with $m^*/m \approx 0.83$ 
is favored by the data above $E_\mathrm{lab} \approx 0.8 A$ GeV
\cite{Li:1989zza,Cassing:1991cy,Fuchs:1996uv,Nara:2019qfd}.
This fact indicates that the values of the optical potential may be 
close to those in the parameter set NS1 at $E_\mathrm{lab} \geq 1$ GeV.
Based on this observation,
the parameter set MD1 is obtained by assuming
values of optical potential similar to those in the NS1 parameter set
at $E_\mathrm{lab} > 1$ GeV
as shown in the dashed line in Fig.~\ref{fig:eosp1}.  Its asymptotic value is 
$U_\mathrm{opt}(\infty)=95$ MeV as indicated in Table~\ref{table:eos}.
The lower panel of Fig.~\ref{fig:eosp1}
compares the energy per nucleon at zero temperature
\begin{equation}
\frac{E}{A} = \frac{e}{\rho_B} - m_N
\end{equation}
as a function of baryon density for different parameter sets.
As is well known, stiffness of the EoS with respect to baryon density
is mainly determined by the value of the effective mass;
smaller effective mass yields stiffer EoS.
Baryon-density dependence in the MD1 parameter set is similar to
the one in the set NS3 which overestimates the flow data,
while NS1, which reproduces the flow data, is softer than NS3
in terms of the baryon density as NS1 has a larger effective mass.
As a comparison, hard EoS ($K=380$ MeV)
from the nonrelativistic Skyrme potential
\begin{equation}
 V_\mathrm{sk}= \alpha \rho_B + \beta \rho_B^\gamma
\end{equation}
is plotted.
The transport models with this Skyrme hard EoS
reproduce the elliptic flow data
at $E_\mathrm{lab}<10 A$ GeV~\cite{Rai:1999hz,Hillmann:2018nmd}.

\begin{figure}[tbh]
\includegraphics[width=\fsize cm]{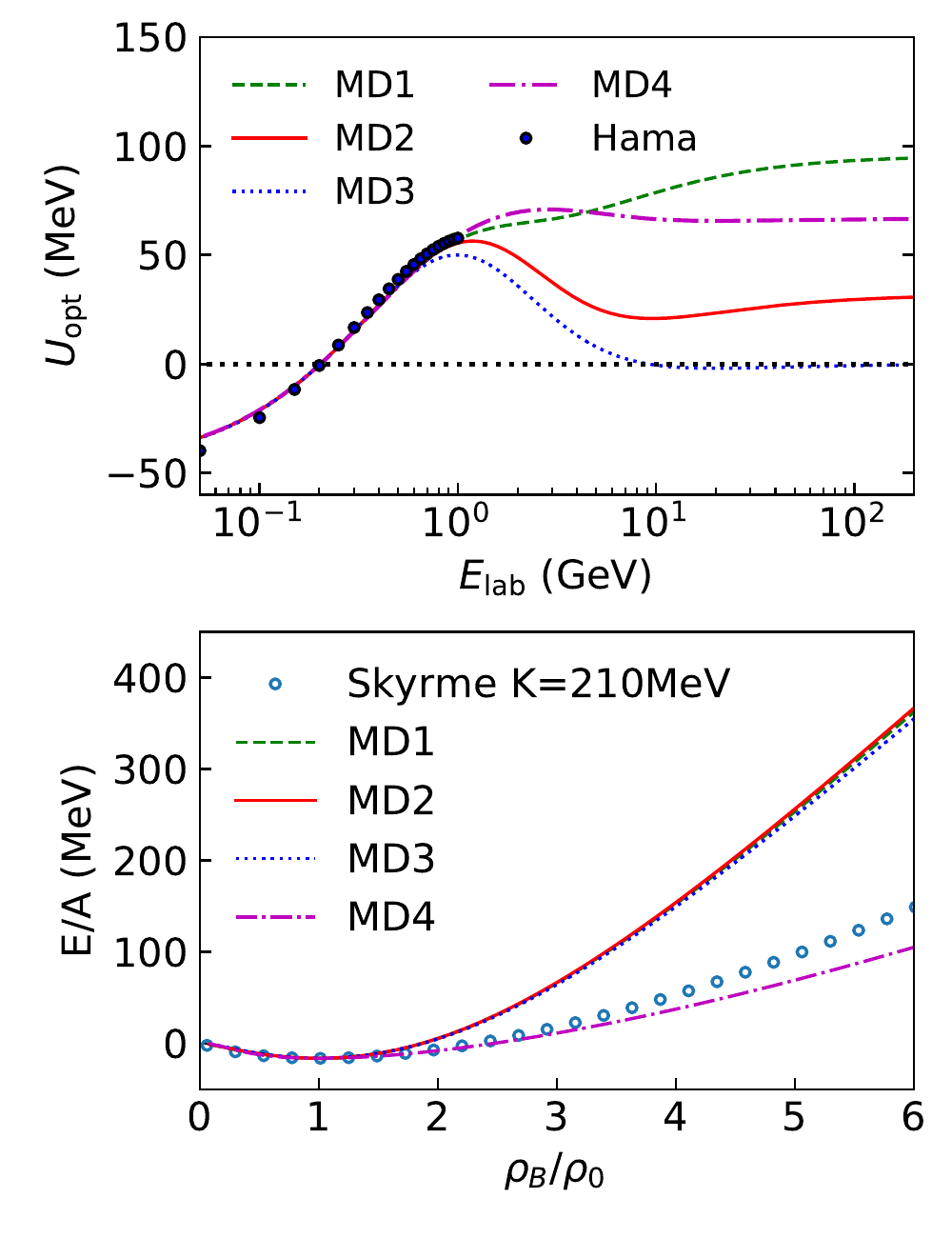}
\caption{Same as Fig.~\ref{fig:eosp1}, but for different parameter sets.
The dashed, solid, dotted, and dot-dashed lines
show the results for
the parameter set MD1, MD2, MD3, and MD4, respectively.
The soft EoS from Skyrme potential with $K=210$ MeV is also shown
by open circles.
}
\label{fig:eosp2}
\end{figure}

There is no experimental information on the nucleon optical potential
at higher energies above $E_\mathrm{lab}=1$ GeV.
Therefore, we consider different values of optical potential
at high energy limit
as shown in the upper panel of Fig.~\ref{fig:eosp2},
keeping the same baryon density dependence  (MD1, MD2, and MD3)
assuming the same effective mass $m^*/m=0.65$ and incompressibility
$K=380$ MeV
(see the lower panel of Fig.~\ref{fig:eosp2}).
We also prepare the soft EoS MD4
($K=210$ MeV and $m^*/m=0.83$),
but its optical potential is almost flat at $E_\mathrm{lab}>1$ GeV.
The baryon density dependence of the MD4 parameter set is as soft as
that of the non-relativistic Skyrme type
potential with $K=210$ MeV as shown
in the lower panel of Fig.~\ref{fig:eosp2}.

\subsection{Relativistic quantum molecular dynamics}

We implement the relativistic EoS constructed above into the
microscopic $N$-body non-equilibrium transport approach
RQMD~\cite{RQMD1989}
which is formulated based on the constraint Hamiltonian
dynamics~\cite{Komar:1978hc}.
The manifestly covariant formulation for the $N$-body dynamics
uses
$8N$ four-vectors $q_i^\mu$ and $p_i^\mu$
$(i=1,\ldots,N)$ for the position and momentum coordinates of particles,
respectively.
Thus, $2N$ constraints are employed to reduce the number of dimensions
from $8N$ to the physical $6N$,
\begin{equation}
  \phi_i \approx 0,~~(i=1,\ldots, 2N),
\end{equation}
where the sign $\approx$ stands for  Dirac's weak equality:
this equality has to be satisfied on the physical $6N$ phase space.
$2N-1$ constraints should be Poincar\'e invariant,
while the $2N$th is not necessarily Poincar\'e invariant,
since it determines the evolution parameter $\tau$.
The Hamiltonian of the $N$-body system is constructed as the linear combination
of $2N-1$ constraints
\begin{equation}
  H = \sum_{j=1}^{2N-1} u_j(\tau) \phi_j
\end{equation}
with the Lagrange multipliers $u_j(\tau)$.
The equations of motion are then given by
\begin{equation}
\begin{aligned}
   \frac{dq_i}{d\tau} &=[H, q_i]
   \approx \sum_{j=1}^{2N-1}u_j\frac{\partial\phi_j}{\partial p_i},\\
   \frac{dp_i}{d\tau} &=[H, p_i]
   \approx -\sum_{j=1}^{2N-1}u_j\frac{\partial\phi_j}{\partial q_i},
   \label{eq:motion}
\end{aligned}
\end{equation}
where the Poisson brackets are defined as
\begin{equation}
   [A, B] = \sum_{k}\left(
    \frac{\partial A}{\partial p_k}
   \cdot \frac{\partial B}{\partial q_k}
  - \frac{\partial A}{\partial q_k}
   \cdot \frac{\partial B}{\partial p_k}
       \right).
\end{equation}
We require that the constraints are conserved in time:
\begin{equation}
  \frac{d\phi_i}{d\tau}=\frac{\partial\phi_i}{\partial\tau}
    + [H, \phi_i] \approx 0.
\end{equation}
As $2N-1$ constraints do not depend explicitly on $\tau$,
the Lagrange multipliers $u_i$ are solved as
\begin{equation}
  u_i \approx -\frac{\partial\phi_{2N}}{\partial\tau}C_{2N,i},
  ~(i=1,\cdots, 2N-1),
\label{eq:constraint}
\end{equation}
where $C_{ij}^{-1}=[\phi_i,\phi_j]$.
In this way,
the equations of motion Eq.~(\ref{eq:motion})
and the Lagrange multipliers Eq.~(\ref{eq:constraint})
uniquely determine
the trajectory of the coupled system of particles in $6N$ phase space.

For our relativistic scalar and vector interaction
in the RQMD.RMF approach,
we choose the $N$ on-mass shell conditions 
\begin{equation}
\phi_i\equiv  p_i^{*2}-m_i^{*2}
       = (p_i - V_i)^2 - (m_i -S_i)^2,~(i=1,\ldots,N)
       \label{eq:onmass}
\end{equation}
for the $i$th particle, where $V^\mu_i$ and $S_i$ are
the single-particle vector and scalar potential.
The remaining $N$ constraints fix the time of $N$ particles.
Here we use the same time fixation constraints proposed
in Ref.~\cite{Maruyama:1996rn,Marty:2012vs},
which equate the all time coordinates
of particles in the reference frame:
\begin{align}
   \phi_{i+N} &\equiv \hat{a}\cdot (q_i - q_N),~~~(i=1,\cdots,N-1),\nonumber\\
   \phi_{2N} &\equiv \hat{a}\cdot  q_N - \tau,
\end{align}
where $\hat{a}$ is a unit-four-vector $\hat{a}=(1,\bm{0})$
in the reference frame~\cite{Maruyama:1996rn}.
A convenient choice may be $\hat{a}=P/\sqrt{P^2}$ with
$P=\sum^N_i p_i$, which equates the time coordinates of all particles
in the overall center-of-mass system~\cite{Marty:2012vs}.

We further make assumption that
the arguments of the potentials are replaced by the free ones%
~\cite{Maruyama:1996rn}.
Then, 
the equations of motion for the $i$th particle are obtained as
\begin{align}
\dot{\bm{x}_i} & =
    \frac{\bm{p}_i^{*}}{p_i^{*0}}
                 +\sum_{j=1}^N\left(
                 \frac{m_j^*}{p_j^{*0}}
                 \frac{\partial m_j^*}{\partial\bm{p}_i}
                 +v^{*}_j \cdot
                   \frac{\partial {V}_{j}}{\partial\bm{p}_i}
                   \right), \nonumber\\
\dot{\bm{p}}_i
          &= -\sum_{j=1}^N\left(
                 \frac{m_j^*}{p_j^{*0}}
                 \frac{\partial m_j^*}{\partial\bm{r}_i}
                 +v^{*}_j \cdot
                   \frac{\partial V_{j}}{\partial\bm{r}_i}
                   \right) ,
\label{eq:eom}
\end{align}
where $v^{*\mu}_i=p^{*\mu}_i/p^{*0}_i$.

Within the RQMD approach,
the scalar density and baryon current
are evaluated by employing the Gaussian wave packet:
\begin{equation}
\rho_{s,i}=\sum_{j\neq i} \frac{m_j}{p_j^0}\rho_{ij},~~
J^\mu_i=\sum_{j\neq i}B_j \frac{p^\mu_j}{p^0_j} \rho_{ij} .
\label{eq:density}
\end{equation}
Here $B_j$ is a baryon number of the $j$th particle. 
Note that we use a free mass $m_j$ and canonical momentum $p_j^\mu$
to compute the scaler density and the baryon current,
since we assumed that the arguments of potentials are replaced by the free ones
in the derivation of the equations of motion.
However, we found that
even though the scalar density and the baryon current
are defined by using effective mass $m_i^*$ and kinetic momentum $p^{*\mu}_i$,
numerical results turn out to be almost unchanged.
The Gaussian $\rho_{ij}$ is given by
\begin{equation}
  \rho_{ij}=\frac{\gamma_{ij}}{(2\pi L)^{3/2}}\exp(q^2_{T,ij}/2L) ,
  \label{eq:gaussian}
\end{equation}
where
$q^2_{T,ij}$ is a distance squared measured in a certain frame,
and $\gamma_{ij}$ is a Lorentz $\gamma$ factor which
ensures the correct normalization of the Gaussian~\cite{Oliinychenko:2015lva}
in Eq.(\ref{eq:gaussian}).
We note that this definition of $\rho_{ij}$ is different from
the so-called interaction density by a factor of two which is defined by
the overlap of density with other Gaussian wave packets
in the QMD approach with Skyrme force.
Throughout this work, the Gaussian width is fixed at $L=1.0$ fm$^2$.
 
There are several choices 
for the reference frame to define a Lorentz invariant distance squared
$q_{Tij}^2$:
\begin{enumerate}[(1)]
\item overall center-of-mass frame,
\item center-of-mass frame between particle $i$ and $j$,
\item rest frame of $j$th particle.
\end{enumerate}
Two-body c.m. frame has been used in most of the RQMD approach,
while overall center-of-mass frame is used
in Ref.~\cite{Marty:2012vs}, which is convenient when simulations are performed
in the overall center-of-mass frame or box simulations,
since the relative distance $q_{T,ij}$
becomes identical to the non-relativistic distance.
This choice would be justified as far as 
the Gaussian width parameter $L$ is less than
the order of the initial Lorentz contraction of the colliding two-nuclei.
The rest frame of particle is employed
in the RLV model~\cite{RLV} to define the distance:
\begin{equation}
  q_{T,ij}= q_{ij} - (q_{ij}\cdot u_j)u_j,
\end{equation}
where $q_{ij}=q_i - q_j$ and $u_j=p_j/m_j$.
For this choice, the $\gamma$ factor in the Gaussian 
becomes $\gamma_{ij} = p^{0}_j/m_j$ which cancels the factor
in the scalar density in Eq.(\ref{eq:density}),
and the scalar density is explicitly Lorentz scalar:
\begin{equation}
\rho_{s,i}=\sum_{j\neq i} \frac{1}{(2\pi L)^{3/2}} 
  \exp\left(
  \frac{q_{ij}^2-(q_{ij}\cdot u_{j})^2}{2L}
  \right).
\end{equation}
We have checked that all of three choices yield 
practically identical results.

In RQMD.RMF with momentum-dependent potential,
the single particle scalar and vector potential
for $i$th particle
are defined as
\begin{equation}
 S_i = \frac{1}{2} g_s \sigma_i + V^\mathrm{MD}_{s,i},~~
 V_{i,\mu} = 
   \frac{B_i}{2}g_v\omega_{i,\mu}
  + B_iV^\mathrm{MD}_{i,\mu}
  \,.
\end{equation}
Here the momentum-dependent potentials are given by
\begin{align}
  V_{s,i}^\mathrm{MD} &= \frac{1}{2}\frac{\bar{g}_s^2}{m_s^2}
       \sum_{i\ne j}^N \frac{m_j}{p^0_j}
       \frac{\rho_{ij}}{1-p_{T,ij}^2/\Lambda_s^2 }\,, \\
  V_{\mu,i}^\mathrm{MD} &= \frac{1}{2}\frac{\bar{g}_v^2}{m_v^2}
       \sum_{i\ne j}^N \frac{p_{\mu,j}}{p^0_j}
       \frac{B_j\rho_{ij}}{1-p_{T,ij}^2/\Lambda_v^2}\,,
  \label{eq:mdqmd}
\end{align}
where $p_{T,ij}$ is a relative momentum between
$i$th and $j$th particle in the two-body center-of-mass frame:
\begin{equation}
p_{T,ij}= p_{ij} -  \frac{(p_{ij}\cdot P_{ij})}{P_{ij}^2} P_{ij}
\end{equation}
where $p_{ij}=p_i-p_j$ and $P_{ij}=p_i+p_j$.

In the actual simulations,
the non-linear $\sigma$-field, as well as the $\omega$-field
at $i$th particle's position,
is evaluated by using a local density approximation~\cite{RBUU,GiBUU,RVUU},
which neglects the derivatives of the scalar and the vector meson field:
\begin{equation}
 m^2_s\sigma_i + g_2\sigma^2_i + g_3\sigma^3_i = g_s \rho_{s,i},~~
 m^2_v \omega_i^\mu = g_v J_i^\mu.
 \label{eq:sigmaomega}
\end{equation}
This approximation is widely applied to
the simulations of high energy nuclear collisions
\cite{Li:1995qm,Larionov:2007hy,Cassing:2015owa}. 
See Ref.~\cite{Weber:1990qd} for the study of the effects of
the meson field radiation and retardation effects within the RBUU approach.

\subsection{Collision term}

Mean-field propagation by the Hamiltonian
is combined with Boltzmann type collision term.
Two-body collision terms are applied by the Monte Carlo method
to simulate particle productions as well as decays by using the transport
code JAM.
Particle productions are modeled by the excitation of
hadronic resonances and strings followed by their decays.
A detailed discussion of the collision term treatment
is found in Refs.~\cite{JAMorg,JAM2}.
In JAM, free cross sections are used in the two-body collisions.
In order to take into account in-medium threshold effects
in two-body collisions,
we evaluate the cross section with 
\begin{equation}
 \sqrt{s_\mathrm{free}}=\sqrt{s^*}- (m_1^* - m_1) - (m_2^*-m_2) \,,
\end{equation}
where $m_i$ and $m^*_i$ are the free and effective hadron mass, respectively,
and $s^*=(p_1^* + p_2^*)^2$,
as employed in the RBUU calculations~\cite{GiBUU}.

Collision term changes the momentum of particles, thus breaks
energy conservation if momentum-dependent potentials are included,
although total momentum is strictly conserved at each collision.
However, the violation of energy conservation
is found to be about 3-5\% level when  momentum-dependent
potentials are included.
We have checked the effects of the energy conservation
by recovering total energy as follows by using the same method in Ref.%
~\cite{Oliinychenko:2016vkg}:
First, go to the center-of-mass frame, then all momenta of the particles
are scaled with the same factor $a$ 
such that
$E_\mathrm{tot} = \sum_i (\sqrt{m_i^{*2} + (a\bm{p}^{*}_i)^2}+V^0_i)$,
where $E_\mathrm{tot}$ is the total energy that we need to recover.
We obtain the factor $a$ by the iteration:
\begin{equation}
a' = \frac{aE_\mathrm{tot}}{\sum_i (\sqrt{m_i^{*2} + (a\bm{p}^{*}_i)^2}+V^0_i)}
\end{equation}
until the desired accuracy is achieved.
It is expected that this procedure has little effect on the flows,
since only the magnitude of momenta is iterated.
Energy conservation is recovered
at each two-body collision in the JQMD model~\cite{Mancusi:2009zz}, which
employs a different algorithm than the one used in JAM
for the treatment of collision term and decay.
In this case, however, we have to update the collision list of
all particles in JAM.
In order to avoid this complication, we recover total energy
at each Hamiltonian time step where the collision list of all particles
has to be updated.
This should be a good approximation,
if time step is small and
the number of collisions or decay is not so large within each time step.
We have checked that, when energy conservation is recovered
until it reaches within 0.1\% accuracy,
we still get the same results.
In order to save computational time, all results in this paper
are obtained without this option.

\section{Results}
\label{sec:results}

In this section, we present the results 
for sideward, directed, and elliptic flow
in midcentral Au + Au and Pb + Pb collisions
from the  RQMD.RMF model
using different equations of state described above.

The centrality cuts in the experimental data usually refer to
cuts in the measured multiplicity distributions.
The centrality cuts in the calculations are usually done
by the analogous cuts in the impact parameter distributions.
In the present paper, directed flow ($v_1$) and elliptic flow ($v_2$)
are analyzed at midcentral collisions, 
which correspond to the impact parameter range
(and multiplicity range) where the $v_1$ flow is rather close to its maximum.
The E895 collaboration~\cite{E895v1,E895v2}
finds that their multiplicity- selected
centrality class corresponds to impact parameters between 5 and 7 fm.
The STAR FXT collaboration~\cite{Wu:2018qih} selects,
at $\sqrt{s_{NN}} = 4.5$ GeV, 
their centrality cut so that it comes close to the E895 one.
Then the impact parameter range 5 to 7 fm corresponds to the 10-25\%
midcentral multiplicity cut in the STAR experiment.
The STAR collaboration uses, at higher energies~\cite{STARv1,STARv2},
a considerably wider multiplicity cut, namely 10-40\%
centrality, which consequently corresponds to a considerably
wider impact parameter range of $4.4 < b < 9.5$ fm.
The NA49 midcentral data\cite{NA49prc} correspond to
an impact parameter range of 5.5  to 9.1 fm.
The FOPI collaboration's centrality cut on their
$v_2$ data~\cite{Andronic:2004cp} is M3, corresponding to
$b =$ 5.5--7.5 fm. However, in Ref.~\cite{Andronic:2004cp},
FOPI shows another cut, $b=$ 7.5--9.5 fm (M4),
with nearly identical $v_2$ values as in the more central M3 cut--hence,
all these experimental observations
suggest a ''midcentral impact parameter range" of $4.6 < b < 9.4$ fm.
This range was previously used to analyze directed flow
by the UrQMD hybrid model collaboration~\cite{Steinheimer:2014pfa}.
This finding suggests using that same ``midcentral" impact parameter range
for calculations at all beam energies, at least up to
$\sqrt{s_{NN}} < 100$ GeV.
We have checked that also a ``midcentral"  cut of 5 to 7 fm yields
nearly the same flow value--hence, the correct results do depend
only weakly on the precise values of the impact parameter cut.

\subsection{directed flow}

\begin{figure}[tbh]
\includegraphics[width=\fsizeb cm]{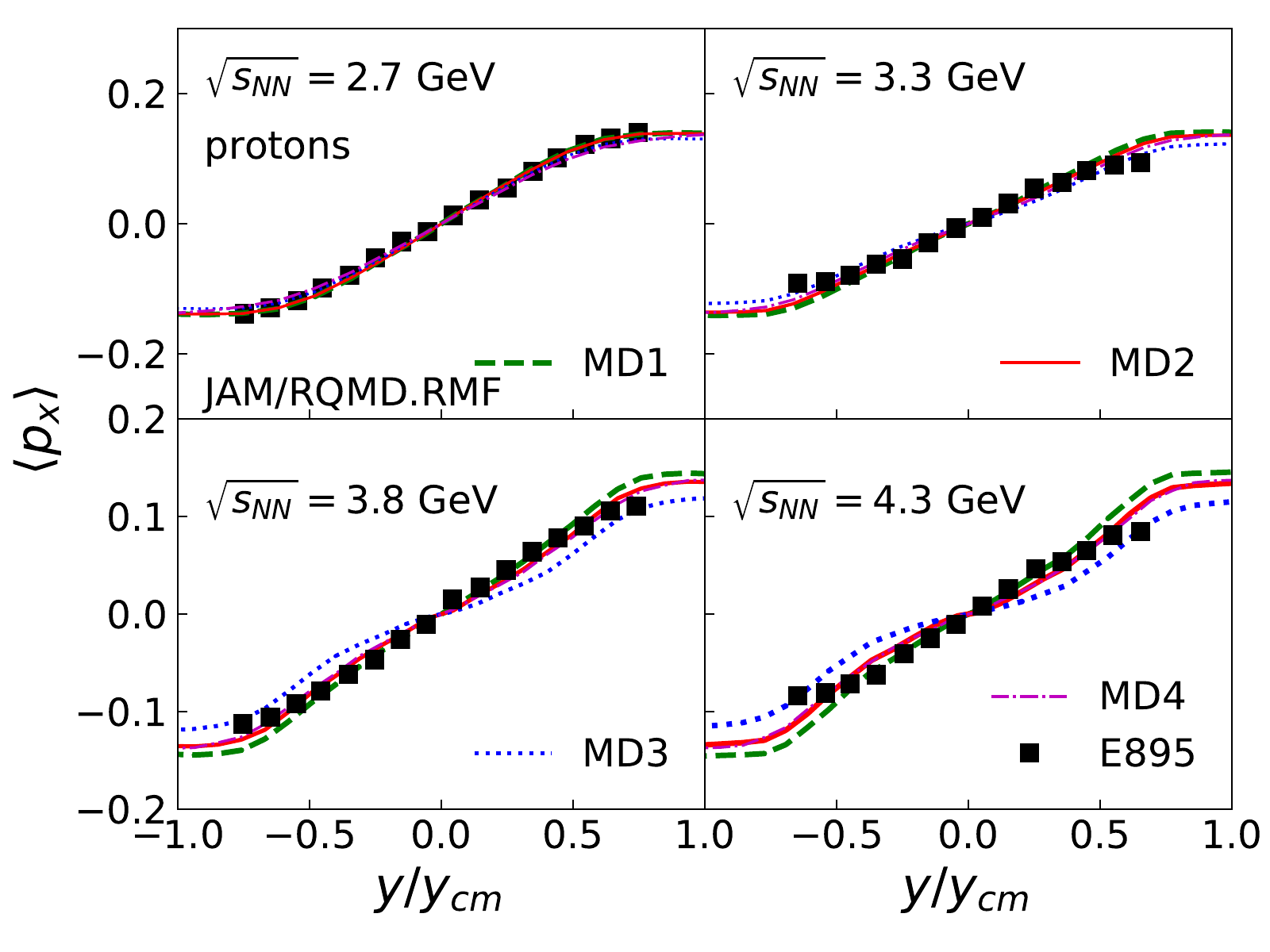}
\caption{Rapidity dependence of proton sideward flow
$\langle p_x\rangle$ in midcentral Au + Au collision
at $\sqrt{s_{NN}}=2.7, 3.3, 3.8$ and 4.3 GeV
($E_\mathrm{lab}=1.85, 4, 6, 8A$ GeV)
from the parameter set MD1
(dashed line), MD2 (solid line), MD3 (dotted line), and
MD4 (dotted-dashed line)
are compared with the E895 experimental data%
~\cite{E895v1}.  
The results of MD4 are not visible because it is nearly identical to the results
of MD2.
}
\label{fig:pxe895}
\end{figure}

Let us first study the optical potential dependence of the
sideward flow $\langle p_x\rangle$
by comparing the parameter sets MD1, MD2, and MD3.
All of them have the same incompressibility $K=380$ MeV
and the effective mass $m^*/m=0.65$, but different strengths
of the optical potential above $E_\mathrm{lab}> 1$ GeV.
Figure~\ref{fig:pxe895} shows the rapidity dependence of the
sideward flow in midcentral Au + Au collisions
at $\sqrt{s_{NN}}=2.7, 3.3, 3.8$ and 4.3 GeV
with different parameter set MD1, MD2, and MD3.
The full squares represent
the experimental data from the E895 collaboration~\cite{E895v1}.
It is seen that all parameter sets yield 
similar results at $\sqrt{s_{NN}}=2.7$ GeV,
since EoS at 2.7 GeV is almost the same among MD1, MD2, and MD3,
due to the constraints from experiments.
As the beam energy increases, the difference among
EoS becomes visible indicating the sensitivity of the
sideward flow to the optical potential.
We note that the results from the set MD4 is identical to the
results from MD2 indicating that directed flow data is insensitive to 
the stiffness of the EoS.

To obtain free protons,
we identify nuclear cluster based on the phase space distribution
of nucleons at the end of the simulation by using a minimum distance chain
procedure; two nucleons are considered to be bound in the same cluster
if the relative distance and momentum between nucleons are less than
4 fm and 0.3 GeV/$c$, respectively.
We have found that the effects of the nuclear cluster as well as
the weak decay of hyperons
on the sideward flow shown in Fig.~\ref{fig:pxe895}
are very small.
However, nuclear cluster effects are large
close to the target and projectile rapidities.

\begin{figure}[tbh]
\includegraphics[width=\fsize cm]{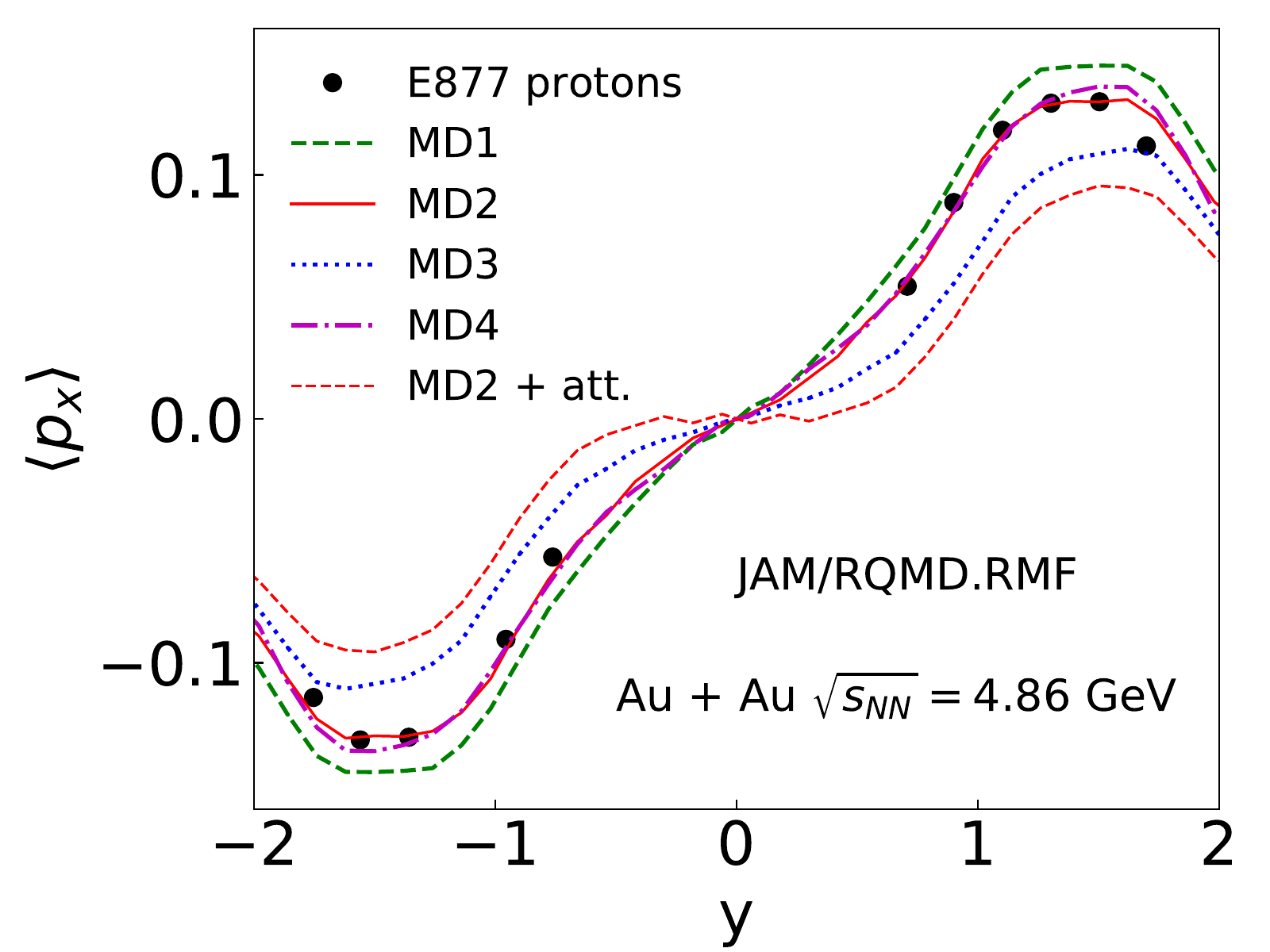}
\caption{Rapidity dependence of proton sideward flow
$\langle p_x\rangle$ in midcentral Au + Au collision
at $\sqrt{s_{NN}}=4.86$ GeV from MD1
(dashed line), MD2 (solid line),
MD3 (dotted line),
MD4 (dotted-dashed line),
and MD2 + attractive orbit
are compared with the E877 experimental data%
~\cite{Barrette:1997pt}.  }
\label{fig:pxe877}
\end{figure}

We expect that the optical potential dependence of the sideward flow
may become significant at higher beam energies.
In Fig.~\ref{fig:pxe877}, 
sideward flow from various EoS are compared for
midcentral Au + Au collision at $\sqrt{s_{NN}}=4.864$ GeV.
Strong sensitivity of the sideward flow to the optical potential is seen.
The MD1 set overestimates the data which has a value
of optical potential similar to that of the NS1 set that has almost flat
optical potential as a function of kinetic energy.
The MD2 set reproduces the data which
has approximately twice smaller optical potential
of $U_\mathrm{opt}\approx 30$ MeV
than that of the set MD1 at $E_\mathrm{lab}=10$ GeV.
However, the set MD4 ($K=210$ MeV, $m^*/m=0.83$)
can also fit the data with
the optical potential $U_\mathrm{opt}\approx$ 70--90 MeV.
Thus, this analysis shows that effective mass parameter
and the optical potential correlate to each other;
smaller effective mass needs smaller optical potential to reproduce
the sideward flow data.
Therefore,  determination of the optical potential
by experiments
should give important constraint on the information
about the properties of excited hadronic matter.

The phase transition to a quark-gluon plasma is
connected to the softening of the EoS, and the signal may be
observed in the directed flow; the slope of proton directed flow
at mid-rapidity becomes negative~\cite{Csernai:1999nf}.
The effects of the softening of the EoS can be efficiently  simulated
by selecting attractive orbit in every two-body scattering
in a microscopic transport simulation~\cite{Nara:2016phs,Nara:2016hbg}.
The result of MD2 + attractive orbit calculation is also
plotted in Fig.~\ref{fig:pxe877}.
The rapidity dependence of the directed flow for this calculation
is similar to the prediction by the hydrodynamic calculation
in Ref.~\cite{Csernai:1999nf}, and the slope at mid-rapidity becomes negative. 
However, experimental data do not indicate
such softening of the EoS at $\sqrt{s_{NN}}=4.86$ GeV.
See Ref.~\cite{E877v1}
for the directed flow at mid-rapidity from the E877 collaboration,
which shows positive slope for protons.

\begin{figure}[tbh]
\includegraphics[width=\fsize cm]{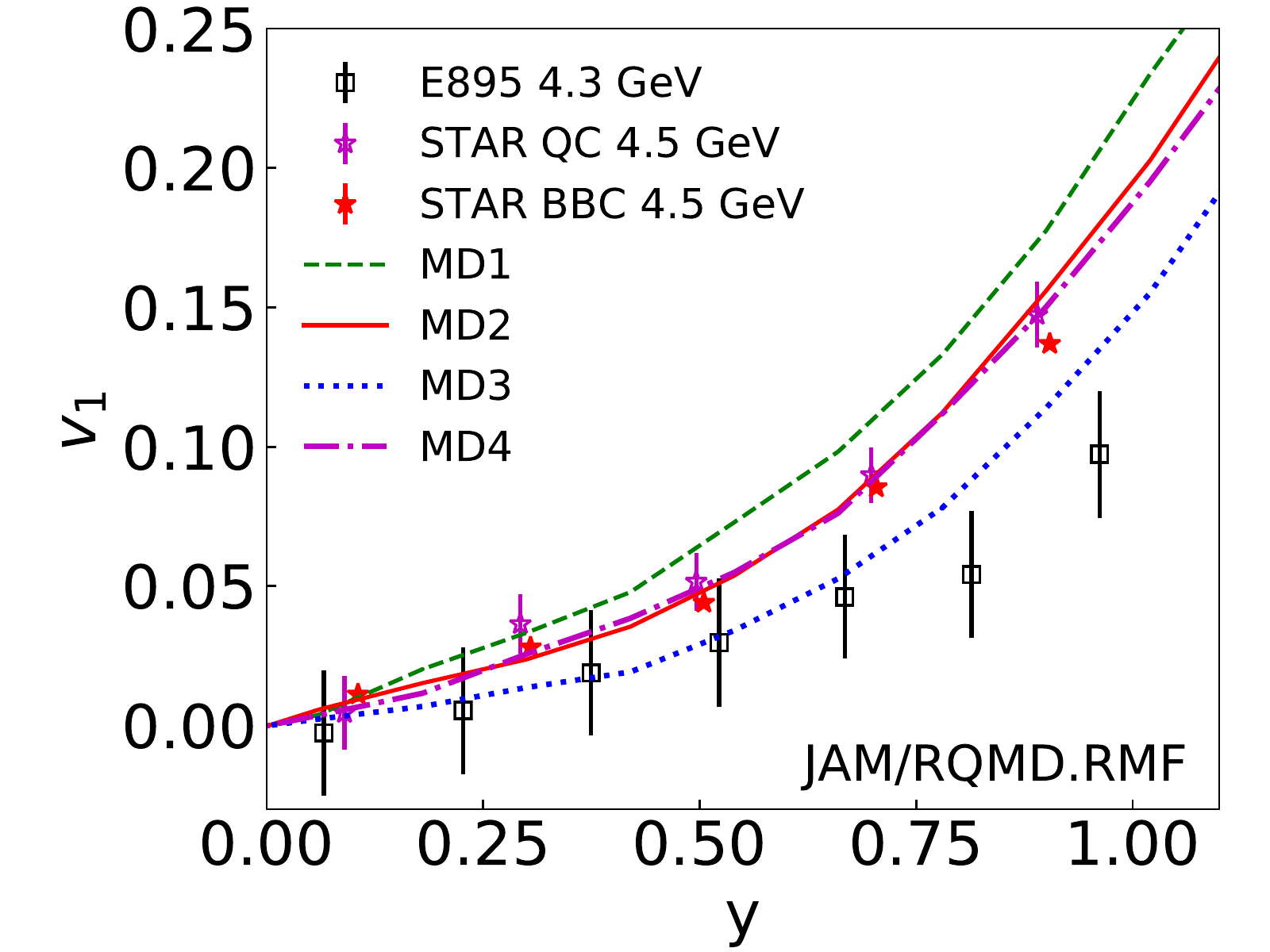}
\caption{Rapidity dependence of proton directed flow $v_1$
in midcentral Au + Au collision
at $\sqrt{s_{NN}}=4.5$ GeV from MD1 (dashed line),
MD2 (solid line), MD3 (dotted line), and MD4 (dotted-dashed line)
are compared with the E895~\cite{E895v1} and STAR data%
~\cite{Wu:2018qih}.
The momentum cut $0.4 < p_T < 2.0$ GeV is imposed, and
only `free' protons are selected in $v_1$ after nuclear coalescence.
}
\label{fig:v1_45}
\end{figure}

Figure~\ref{fig:v1_45} shows the rapidity dependence of the directed flow
$v_1 = \langle p_x / p_T \rangle$
in midcentral Au + Au collisions at $\sqrt{s_{NN}}=4.5$ GeV
from RQMD.RMF simulations with different parameter sets, which are
compared with the STAR preliminary data~\cite{Wu:2018qih}
and E895 data~\cite{E895v1}.
It is also seen that directed flow is sensitive to the EoS.
The results of the MD2 and MD4 parameter set,
which fit the E895 sideward flow data,
are consistent with the STAR data.
However, MD2 and MD4 overestimate the E895 directed flow data.
We note that this discrepancy has been already discussed within
three-fluid dynamics (3FD) simulations~\cite{Ivanov:2016sqy}.
3FD reproduces the sideward flow $\langle p_x \rangle$
at AGS energies, while the agreement of the calculated directed
flow $v_1$ with the E895 data is worse than the calculated sideward flow.
It seems that STAR FXT data may clarify the inconsistency of the old AGS data.

\begin{figure}[tbh]
\includegraphics[width=\fsizeb cm]{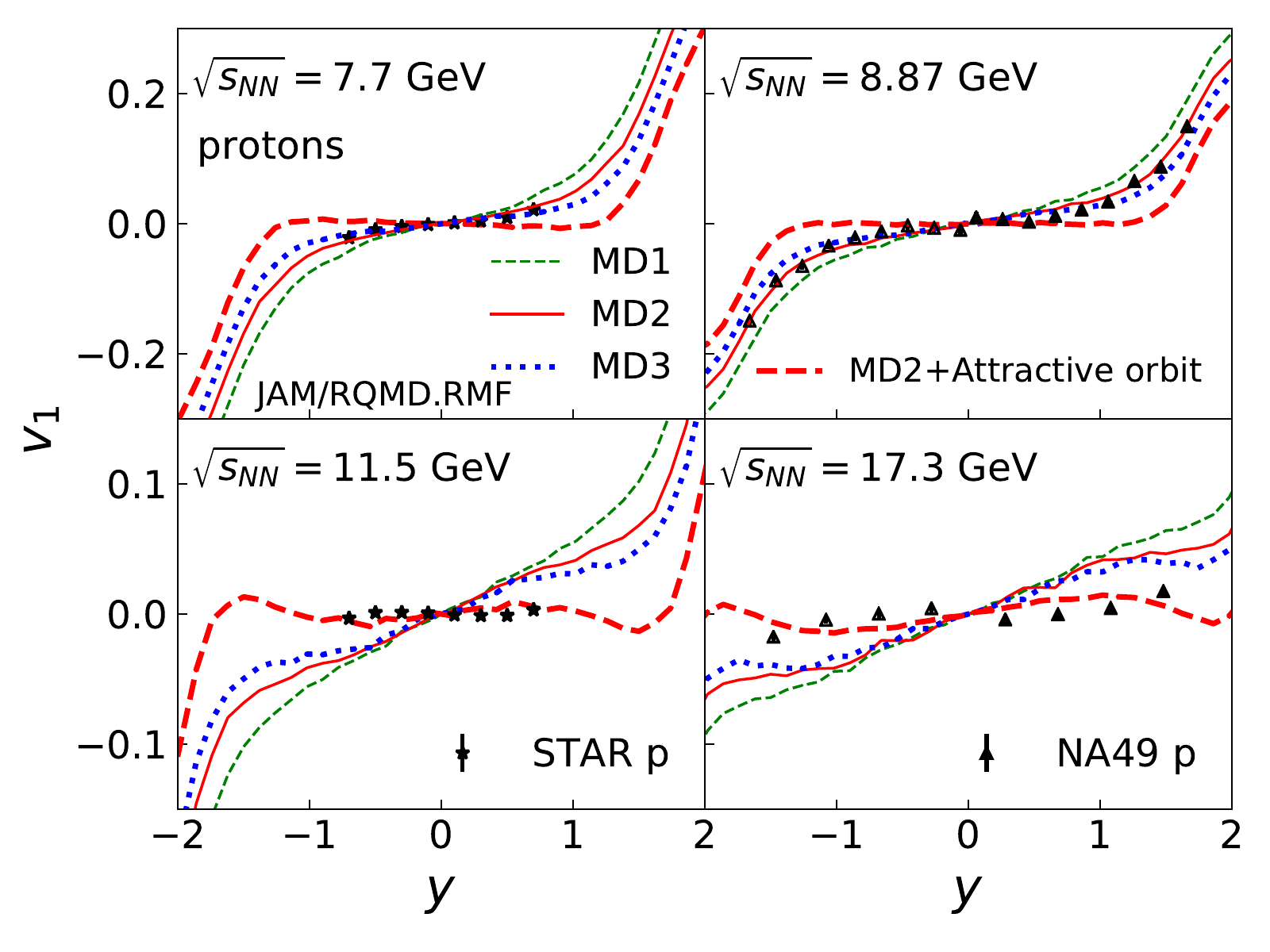}
\caption{Rapidity dependence of proton directed flow
$v_1$ in midcentral Au + Au 
at $\sqrt{s_{NN}}=7.7, 11.5$
and Pb + Pb collisions at 8.87, 17.3 GeV
from MD1
(dashed line), MD2 (solid line), MD3 (dotted line),
and MD2 + attractive orbit simulations (thick dashed line)
are compared with the NA49~\cite{NA49prc} and STAR experimental data%
~\cite{STARv1}.  
}
\label{fig:v1sps}
\end{figure}

Let us look at the directed flow at higher beam energies.
Figure~\ref{fig:v1sps} compares the rapidity dependence of the proton
directed flow from the RQMD.RMF calculations 
in midcentral Au + Au at $\sqrt{s_{NN}}=7.7, 11.5$ GeV
and Pb + Pb at $\sqrt{s_{NN}}=8.87, 17.3$ GeV
with STAR~\cite{STARv1} and NA49 data~\cite{NA49prc}.
The RQMD.RMF result with the EoS parameter set MD2 is in good agreement
with the data at $\sqrt{s_{NN}}=7.7$ GeV,
while its slope at midrapidity is slightly higher than that of data
at $\sqrt{s_{NN}}=8.87$ GeV, which indicates the onset of
the softening of the EoS.
This is clearly seen at $\sqrt{s_{NN}}=11.5$ and 17.3 GeV,
where all calculations with the `normal' EoS 
predict strong positive slope in contrast to the data which
show negative slope.
We note that the rapidity dependence of the directed flow
from the MD4 parameter set is the same as that of the MD2 parameter set.
 
We also compare the results from the attractive orbit simulation
with the MD2 EoS
which mimics a softening of the EoS~\cite{Nara:2016phs,Nara:2016hbg}.
As shown in Ref.~\cite{Nara:2016phs},
the pressure due to the attractive orbit is reduced as low as the one
in a typical EoS with a first-order phase transition.
It is seen that the attractive orbit simulations predict significant reduction
of the directed flow slope, and their results are close to the data
at $\sqrt{s_{NN}}=11.5$ and 17.3 GeV. On the other hand,
it is inconsistent with the data at $\sqrt{s_{NN}}=7.7$ GeV.
Note that the proton negative flow at higher energies $\sqrt{s_{NN}}>30$ GeV,
where secondary interactions start after two nuclei pass through each other,
can be understood by the geometrical effects%
~\cite{Snellings:1999bt,Zhang:2018wlk}.
It is very important to notice, however, that
this geometrical interpolation is not applicable at $\sqrt{s_{NN}}<30$ GeV,
because secondary hadronic interactions
alter the dynamics during the overlapping times
of the colliding nuclei
as hadronization time is less than the crossing time~\cite{Sorge:1997nv}.
Thus, negative proton slope at $\sqrt{s_{NN}}<20$ GeV
cannot be explained by the geometrical effects.
Therefore, our analysis support that the collapse of the directed flow
around $\sqrt{s_{NN}}\approx 10$ GeV
discovered by the experiments is an evidence of the softening of the EoS.
We note that
the directed flow data is in favour of the crossover EoS
within the 3FD calculations%
~\cite{Konchakovski:2014gda,Ivanov:2014ioa,Ivanov:2016sqy}.

\subsection{Elliptic flow}

We now examine the beam energy dependence of the elliptic flow
$v_2 = \langle \cos 2\phi \rangle$.
%
\begin{figure}[tbh]
\includegraphics[width=\fsize cm]{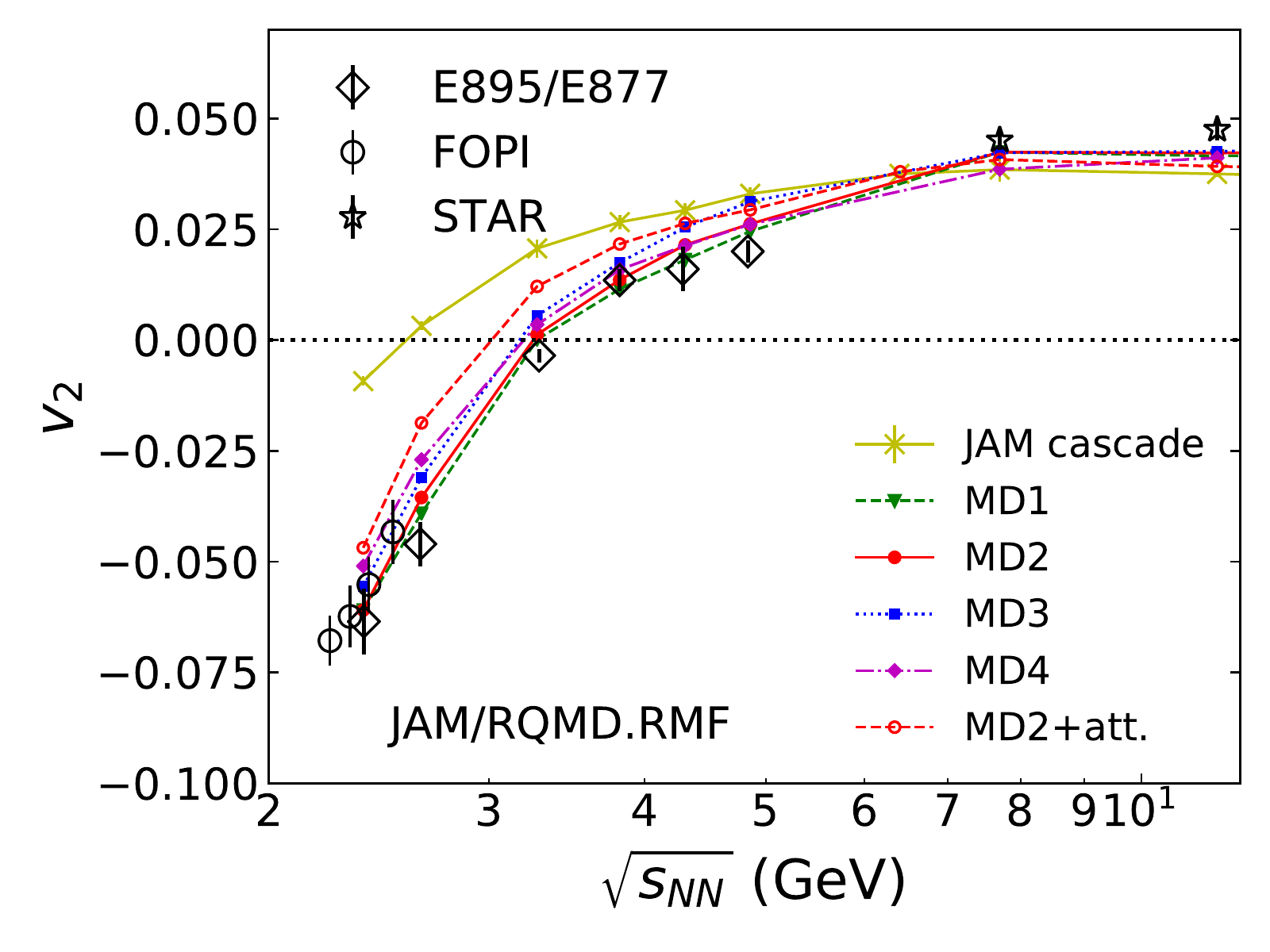}
\caption{Beam energy dependence of
elliptic flow $v_2$ of
proton ($\sqrt{s_{NN}}<5$ GeV)
and charged hadron ($\sqrt{s_{NN}}> 5$ GeV)
in midcentral Au + Au collision
from MD1 (triangles)
, MD2 (circles), MD3 (squares), MD4 (diamond),
and MD2 + attractive orbit (open circles)
are compared with the experimental data
FOPI~\cite{Andronic:2004cp},
E895/E877 \cite{E895v2} and, STAR~\cite{STARv2}.
The STAR data~\cite{STARv2} for $v_2$ are for charged hadrons.
}
\label{fig:v2e895}
\end{figure}
%
Figure~\ref{fig:v2e895} displays
the beam energy dependence of the elliptic flow at midrapidity
in midcentral Au + Au collisions
at $\sqrt{s_{NN}}<5$ for protons,
and at $\sqrt{s_{NN}}>6$ GeV for charged hadrons.
Experimental data for the elliptic flow
is consistent with the strong repulsive interactions in RQMD.RMF
with the MD1 and MD2 parameter set,
which generate strong out-of-plane emission (squeeze-out),
while the parameter set MD3 which has weak optical potential
predicts less out-of-plane emission.
The soft EoS MD4
generates weaker elliptic flow at lower beam energies.
Thus, elliptic flow data 
exclude very weak optical potential and  a soft EoS at low energies.

If there is a first-order phase transition, out-of-plane emission
is suppressed, and enhancement of $v_2$ and $v_4$ is predicted 
within the cascade model with modified scattering style
\cite{Nara:2017qcg,Nara:2018ijw}.
To see such softening effects in our approach, we plot
in Fig.~\ref{fig:v2e895} the results from MD2 parameter set
with attractive orbit simulation.
MD2 with attractive orbit simulations yield less out-of-plane emission
at AGS energies, while they do not change the elliptic flow much
at SPS energies.
Experimental data for elliptic flow, however, do not support the softening
of the EoS at AGS energies.
There is no  data between $5 <\sqrt{s_{NN}}<7.7$ GeV.
A new data at $\sqrt{s_{NN}}\approx 6$ GeV should
provide further confirmation 
about the EoS.

Our hadronic approach predicts less elliptic flow
at $\sqrt{s_{NN}}=11.5$ GeV.
The elliptic flow at higher energies increases due to strong
in-plane emission, which is consistent with hydrodynamical calculations%
~\cite{Karpenko:2015xea,Shen:2020},
and microscopic transport models with partonic phase%
~\cite{Lin:2001zk,Konchakovski:2011qa,Konchakovski:2012yg}.

\section{Summary}
\label{sec:summary}

We have extended a relativistic quantum molecular dynamics model based on
the relativistic mean-field theory
by including momentum-dependent potentials.
This approach has been implemented into the JAM transport code to study
the heavy-ion collisions at high baryon density region.
We found that the directed flow is very sensitive to the optical potential,
and there is a correlation between the effective mass parameter
at the normal nuclear matter density and the strength of the optical potential.
Namely, smaller effective mass requires smaller optical potential
to reproduce the directed flow data at $3 <  \sqrt{s_{NN}} < 8$ GeV.
Thus experimental information on the optical potential from $pA$ collisions
at these energy ranges will allow us to constrain EoS.
 
It is also shown that the beam energy dependence of the elliptic flow
at mid-rapidity is well described by the same parameter set which reproduces
the directed flow data.
On the other hand,
stiff EoS is required to describe
the strong squeeze-out at lower beam energies of $\sqrt{s_{NN}}<3$ GeV,
which is consistent with the transport calculations
within the Skyrme type potential in Refs.%
~\cite{Danielewicz:2002pu,Hillmann:2018nmd,E895v2,
Larionov:2000cu,Andronic:2004cp}
and the RBUU calculations~\cite{Maruyama:1993jb,Sahu:1998vz}.
Within the non-relativistic QMD models with
the non-relativistic Skyrme potentials, the kaon production
at the Bevalac and at GSI's SiS18  was studied
by Fuchs, Aichelin, and Hartnack, \textit{et al}.
~\cite{Sturm:2000dm,Fuchs:2000kp,Hartnack:2005tr,Hartnack:2011cn}
They suggest that soft EoS with $K=210$ MeV can be extracted
from the analysis of the data at beam energies below $E_\mathrm{lab} = 2$ AGeV. 
Also, $v_1$ and $v_2$ are consistent--within the non-relativistic IQMD model
with a non-relativistic soft EoS~\cite{FOPI:2011aa,Fevre:2015fza}.
Hence, the claim is that the EoS extracted from the data
at these moderate energies, $E_\mathrm{lab} < 2$ AGeV, is soft;
with $K=210$ MeV, independent of
input parameters which are not precisely known.
Our results are quantitatively not consistent with
these IQMD results
~\cite{Sturm:2000dm,Fuchs:2000kp,Hartnack:2005tr,FOPI:2011aa,Hartnack:2011cn,Fevre:2015fza}.
The present paper, which does not investigate kaon yields
at $E_\mathrm{lab} <2$ AGeV, focusses on baryon flow.
Standard free-space elastic and inelastic scattering
cross sections are used, without in-medium modifications,
and without change of meson properties.
Non-relativistic codes do not have different relativistic transformation laws
for scalar effective mass attraction and vector repulsion.
Hence, this systematic difference may the origin for the difference
between non-relativistic IQMD
and relativistic codes RQMD.RMF.

Our approach reproduces the directed and the elliptic flow data
at $2.3 < \sqrt{s_{NN}} < 8$ GeV simultaneously with the parameter set MD2.
In contrast, this approach does not describe the collapse of the proton directed
flow at $8 < \sqrt{s_{NN}} < 20$ GeV  unless taking into account
the effects of a softening of the EoS.
We simulate effectively a softening of the EoS
by imposing attractive orbit
at each two-body collision. This method provides a good description
of the directed flow at $\sqrt{s_{NN}}> 8$ GeV.
However, we still cannot explain the beam energy dependence of
the directed flow in a single consistent framework.
We note that most of the theoretical calculations
predict the collapse of the directed flow below $\sqrt{s_{NN}}\approx 6$ GeV.
Thus, it remains to be understood why softening is seen at
$10 < \sqrt{s_{NN}} < 20$ GeV, if the collapse of the directed flow
is certainly due to the softening of the EoS.
Furthermore,
it is still premature to make an unambiguous conclusion that
the collapse of the directed flow is a signature of
a first-order phase transition.
Theoretically, mean-field approach is a favored method to study EoS dependence
in a dynamical simulation.
As a future work, it would be interesting to simulate chiral phase transition
within the RQMD approach based on chiral mean field model%
~\cite{Motornenko:2019arp}.

\begin{acknowledgments}
We thank W. Cassing, C. Hartnack and J. Steinheimer
for valuable comments on the manuscript,
and D.Keane and Y. Wu for clarifying the STAR and E896 data.
This work was supported in part by the
Grants-in-Aid for Scientific Research
from JSPS (JP17K05448 and JP19K03833).
%
%
H. S. thanks the Walter Greiner Gesellschaft
for the Judah M. Eisenberg Laureatus Professur.
Computational resources have been provided by GSI, Darmstadt, and LOEWE CSC, 
Goethe Universit\"at Frankfurt.
\end{acknowledgments}

\begin{appendix}
\section{Equation of motion}

In RQMD, scalar density and vector potential are computed
by using the Gaussian $\rho_{ij}$:
\begin{equation}
  \rho_{s_i} 
   =\sum_{j\neq i} f_j \rho_{ij},~~~
  V^\mu_i = C_v B_i \sum_{j\neq i} B_j u^{\mu}_j \rho_{ij}
\end{equation}
where $f_j=\dfrac{m_j}{p^{0}_j}$,
$u^{\mu}_{j} =  \dfrac{p^{\mu}_j}{p^0_j}$,
and $B_j$ is a baryon number of particle $j$.
The Gaussian $\rho_{ij}$ is given by
\begin{equation}
 \rho_{ij} = \frac{\gamma_{ij}}{(2\pi L)^{3/2}}
 \exp \left[ \frac{q_{T,ij}^2}{2L} \right] .
\end{equation}
The equations of motion Eq.~(\ref{eq:eom}) can be computed as
\begin{align}
\bm{\dot{r}}_i
  &= \frac{\bm{p}_i^{*}}{p_i^{*0}}
 +\sum_{j\neq i} \Big[
  D_{ij}\frac{\partial\rho_{ij}}{\partial\bm{p}_i}
 +D_{ji}\frac{\partial\rho_{ji}}{\partial\bm{p}_i}
  \nonumber\\
& ~~~~~~~~~~~~ +\left(D_j \frac{\partial f_i}{\partial\bm{p}_i}
+A_j^\mu \frac{\partial u_{i\mu}}{\partial\bm{p}_i}\right)\rho_{ji}
   \Big]
    \\
\bm{\dot{p}}_i &= -\sum_{j\neq i}
 \left[
  D_{ij}\frac{\partial\rho_{ij}}{\partial\bm{r}_i}
 +D_{ji}\frac{\partial\rho_{ji}}{\partial\bm{r}_i}
  \right]
\end{align}
where
\begin{align}
D_{ij}&=D_{i}f_j + A_{ij}^\mu u_{j\mu},\\
D_{i} &= \frac{m_i^*}{p^{*0}_i}\frac{\partial S_i}{\partial\rho_{si}},\\
A_{ij}^\mu &= 
	   C_v B_iB_j v_i^{*\mu} \,.
\end{align}
When the two-body or overall center-of-mass frame is used to define
$q_{T,ij}$, the Gaussian $\rho_{ij}$ is symmetric: $\rho_{ij}=\rho_{ji}$.
For the momentum-dependent potentials, we need derivatives of
an additional terms:
\begin{equation}
  \bar{\rho}_{ij} = D(\bm{p}_{ij}^2)\rho_{ij},~~~
  D(\bm{p}_{ij}^2) =  \frac{\bar{C}}{1+\bm{p}_{ij}^2/\Lambda^2}.
\end{equation}
In the case of the nonlinear $\sigma$ field
\begin{equation}
  m_\sigma^2 \sigma_i + g_2\sigma_i^2 + g_3\sigma_i^3 = g_s\rho_{si}
\end{equation}
The derivatives
$\partial S_i/\partial\rho_{si}$
can be obtained by
\begin{equation}
\frac{\partial S_i}{\partial\rho_{si}}
=-g_s\frac{\partial\sigma_i}{\partial\rho_{si}}
=\frac{-g_s^2}{m_\sigma^2 + 2g_2\sigma_i + 3g_3\sigma_i^2}
\end{equation}

\end{appendix}

\end{document}